\documentclass[oneside,a4paper]{article}

\usepackage{graphicx}%
\usepackage{multirow}%
\usepackage{amsmath,amssymb,amsfonts}%
\usepackage{amsthm}%
\usepackage{mathrsfs}%
\usepackage[title]{appendix}%
\usepackage{xcolor}%
\usepackage{textcomp}%
\usepackage{manyfoot}%
\usepackage{booktabs}%
\usepackage{algorithm}%
\usepackage{algorithmicx}%
\usepackage{algpseudocode}%
\usepackage{listings}%
\usepackage{lineno}
\usepackage{acronym}
\usepackage{subcaption}

\usepackage[numbers]{natbib}
\newcommand{\eq}[1]{
\begin{eqnarray}
	#1
\end{eqnarray}
}
\newcommand{\Sch}{Schr\"odinger }

\newcommand{\CP}{\mbox{\scriptsize CP}}
\newcommand{\cp}{\mbox{\tiny CP}}
\newcommand{\JCP}{J_{\CP}}
\newcommand{\Jcp}{J_{\cp}}
\newcommand{\dCP}{\delta_{\CP}}

\newcommand{\commentout}[1]{}

 \graphicspath{ {./images/} } 

\acrodef{MSW}{Mikheyev-Smirnov-Wolfenstein}
\acrodef{SM}{Standard Model}
\acrodef{BSM}{Beyond Standard Model}
\acrodef{d.o.f.}{degree of freedom}
\acrodef{CP}{Charge-Parity}
\acrodef{PMNS}{Pontecorvo–Maki–Nakagawa–Sakata}
\acrodef{LUT}{look-up table}
\acrodef{DUNE}{Deep Underground Neutrino Experiment}
\acrodef{CJM}{Cyclic Jacobi Method}
\acrodef{NH}{Normal Hierarchy}
\acrodef{IH}{Inverted Hierarchy}
\acrodef{NSNI}{Non-Standard Neutrino Interactions}
\acrodef{SDS}{Sequential Diagonalization Strategy}
\acrodef{SJD}{Sequential Jacobi Diagonalization}

\begin{document}
\title{The case for adopting the sequential Jacobi’s diagonalization algorithm in neutrino oscillation physics}


\author{Gustavo A. Valdiviesso}
\date{\today}

\maketitle

\abstract{Neutrino flavor oscillations and conversion in an interacting background (MSW effects) may reveal the charge-parity violation in the next generation of neutrino experiments. The usual approach for studying these effects is to numerically integrate the Schrödinger equation, recovering the neutrino mixing matrix and its parameters from the solution. This work suggests using the classical Jacobi's diagonalization in combination with a reordering procedure to produce a new algorithm, the Sequential Jacobi Diagonalization. This strategy separates linear algebra operations from numerical integration, allowing physicists to study how the oscillation parameters are affected by adiabatic MSW effects in a more efficient way. The mixing matrices at every point of a given parameter space can be stored for speeding up other calculations, such as model fitting and Monte Carlo productions. This approach has two major computation advantages, namely: being trivially parallelizable, making it a suitable choice for concurrent computation, and allowing for quasi-model-independent solutions that simplify Beyond Standard Model searches. 
}

\section{Introduction}\label{sec:Intro}

Neutrino flavor oscillations in the presence of matter are described by a continuously varying, finite-dimension set of \Sch equations along a propagation path, in what is know as the \ac{MSW} effect\cite{Wolfenstein1978,MS1985}. The varying nature of such backgrounds prevents any practical case from being solved analytically, with numerical methods being the only option. Although these computations by themselves are not intensive, the problem scales in complexity when dealing with any sort of model fitting, or any situation requiring the system to be solved for a large number of configurations. Moreover, knowing the values of the oscillation parameters, mixing angles and mass differences, as a function of a model's parameter space gives valuable insight into neutrino physics itself. In order to map the oscillation parameters, the solutions are used to find the eigenvalues and eigenvectors of the mixing matrix, which adds a second numerical task to the computational load. This second step has a inherent complication: by definition, linear algebra algorithms are agnostic to eigenvalues and eigenvectors ordering. In fact, the output of a numerical diagonalization algorithm has unpredictable ordering, forming what is know as a Newton's Fractal\cite{hubbard2001}. However, this ordering has implication for neutrino physics. The \ac{PMNS} parametrization assumes a mass/flavor ordering when defining the mixing matrix in vacuum and, this ordering, has to be known when matter effects are present in order to properly recover the oscillation parameters. This work makes the case for the use of Jacobi's diagonalization algorithm\cite{Jacobi1846, Eberlein1962, Hansen1963}, which finds both the set of eigenvalues and eigenvectors at the same time. This is accomplished by swapping the diagonalizing with the integration steps, finding the eigenvectors before solving the \Sch equation. This allows us to find the correct ordering of its eigenvalues by a simple comparison to neighbours, maintaining the \ac{PMNS} parametrization which connects the matter-affected oscillation parameters to their vacuum counterparts. By storing the eigenvectors in a \ac{LUT}, the strategies proposed here can trivially offload the computation of fitting and mapping tasks. This procedure is particularly advantageous for exploring exotic matter backgrounds and flavor-changing \ac{BSM} interactions, which are common avenues of study in the neutrino research field. Since future experiments such as the \ac{DUNE}\cite{DUNE2020} will be able to explore \ac{BSM} physics, new algorithms with better computational complexity are required. 

 The modern version of the Jacobi method for the diagonalization of Hermitian matrices is reviewed, followed up by the addition of extra steps with the goal of preserving any preexisting parametrization. This is the \ac{SDS}, which is accomplished by successively comparing the eigenvalues over an arbitrary smooth path in small, discrete steps, transporting the ordering of the eigenvector across the parameter space. This algorithm offers advantages to neutrino physicists, mainly when studying adiabatic evolution in active interacting media. See Refs. \cite{Blennow2013, Maltoni2016} and \cite{Smirnov2016} for excellent reviews. Although analytical solutions do exist for the most common scenarios\cite{Zaglauer1988,Kimura2002}, these are \ac{SM} dependent and cannot be easily expanded for more general models, such as the study of \ac{NSNI} and sterile neutrinos.  

Although the methods described here are aimed at neutrino physics, this paper is organized in such a way that the methodology can be appreciated by a more general reader. Section \ref{sec:Param} defines the ordering problem and its implications; Section \ref{sec:Jacobi} reviews the original Jacobi algorithm for a Hermitian matrix; Sec. \ref{sec:SeqDiag} outlines the \ac{SDS}, which ensures parametrization over a continuous path; And finally, Section \ref{sec:Neutrinos} illustrates an example application in neutrino physics using an example which can also be compared with an analytical solutions, followed by Section \ref{sec:Conc} with the conclusions. After this, two appendices showcase discussions and details that might not be of interest for the general reader: \ref{sec:Test} performs a benchmark test by solving a random case and comparing it with its analytical solution; and finally, \ref{sec:Eff} contains an analysis of convergence, precision and stability of the algorithm. In the remainder of this text, the words \emph{parametrization} and \emph{ordering} are used interchangeably since, in the context of neutrino oscillations, one implies the other.
\bigskip

\section{Parametrized Hermitian Matrices}\label{sec:Param}

Consider a Hermitian matrix $A$ of order $n$, with $n(n-1)/2$ independent elements $A_{jk}\in\mathbb{C}$. The developing a physical model one may want to describe each element as continuous function over a $p$-dimensional parameter space,  i.e., $A_{jk}\equiv A_{jk}(\vec{q})$, with  $\{ \vec{q}=(q_1,\ldots,q_p)\, |\, q_i\in\mathbb{R}\}$. In this situation, its real eigenvalues $\lambda_k$ and corresponding eigenvectors $V_k$ are also functions of $\vec{q}$. By the spectral theorem, all Hermitian matrices are normal matrices and, as such, can be written as $A=U\ D\ U^\dagger$, where $D=\mbox{diag}\left( \lambda_1,\lambda_2,\ldots,\lambda_n\right)$ and 

\eq{\label{eq.U_Definition}
U =
\left[\begin{array}{cccc}
\left(\!\!\begin{array}{c} \vdots \\ V_1 \\ \vdots  \\ \end{array}\!\!\right)
\left(\!\!\begin{array}{c} \vdots \\ V_2 \\ \vdots  \\ \end{array}\!\!\right)
\begin{array}{c} \\ \cdots \\ \\ \end{array}
\left(\!\!\begin{array}{c} \vdots \\ V_n \\ \vdots  \\ \end{array}\!\!\right)
\end{array}\right]
}

\noindent where $U$ is a unitary transformation, i.e., $U\ U^\dagger=1$ and. The ordering of the $\lambda_k$ in $D$ might be a arbitrary, but is assumed to have physical meaning and so it has to be preserved. Note that, by choosing to represent the eigenvalues as the elements of $D$ and the eigenvector as column of $U$, their ordering is preserved in these matrices, by definition, and the relation $A\ V_k = \lambda_k A$ becomes equivalent to $A\ U = U\ D$. For the sake of brevity the pair $\{D,U\}$ will be referred to as the \emph{eigensystem} of $A$. 

We are interested in describing the eigensystem of Hermitian matrices as a function of the original parametrization $\vec{q}$, i.e, since $A\equiv A(\vec{q})$, so must be $D\equiv D(\vec{q})$ and $U\equiv U(\vec{q})$. The Jacobi diagonalization method is the well suited for this goal since it results in the complete eigensystem, already in the form $\{D,U\}$. However, the resulting ordering of their columns is never guaranteed since this information is arbitrary and not related to the method's input, $A$. In fact, the resulting ordering is unpredictable and its dependence on the elements $A_k$ are fractal-like, presenting self-similarity and recursive characteristics\cite{hubbard2001}. In practice, this means that by defining the first element of $D(\vec{q})$ as $\lambda_1(\vec{q})$, one should not assume that the first element of  $D(\vec{q}+\Delta\vec{q})$ is still $\lambda_1(\vec{q}+\Delta\vec{q})$, even after an arbitrarily small step $\Delta\vec{q}$. This implication prevents us from reconstruction the functions $\lambda_k(\vec{q})$ and $V_k(\vec{q})$, unless we transport the ordering information along, with every step. This strategy will be addressed in Sec.\ref{sec:SeqDiag}, after the following review of the Jacobi method.
\bigskip

\section{Jacobi's Algorithm}\label{sec:Jacobi}

The original algorithm proposed by Carl G. J. Jacobi in 1845\cite{Jacobi1846} established a numerical procedure to calculate eigenvalues of a real, symmetric matrix. Since then, several variations were developed in the literature, including an extension to a general complex matrix\cite{Eberlein1962,Hansen1963}. The focus of this work is the diagonalization of Hermitian matrices, which is reviewed in this section for completion sake, using notation and steps mainly based on Ref. \cite{NumRecipesCpp}. For a more comprehensive review, see Ref. \cite{Golub2000}.

Lets start by defining a way to measure the magnitude of a matrix's off-diagonal elements, $d^2$ given by

\eq{\label{eq:d2}
d^2=\frac{2}{n(n-1)}\sum_{i>j}^n A_{ij}\, A_{ji}\,,
}

\noindent Jacobi proved that for all Hermitian matrices $A$ there is an infinite sequence $A_0$, $A_1$, $\cdots$ $A_k$, $A_{k+1}$ $\cdots$, with off-diagonal magnitudes $d^2_{k+1}<d^2_k$ for any $k$, meaning that $A_k$ converges to a diagonal matrix as $k\rightarrow 0$. The sequence in Eq.$\ref{eq:d2}$ has general term given by

\eq{
A_{k+1}&=&S_{k}^\dagger\,A_{k}\,S_{k}\,\,, \hspace{0.5cm}\mbox{or} \label{eq:Ak}\\
&=& S_{k}^\dagger S_{k-1}^\dagger\cdots S_{0}^\dagger\,\, A_0\,\, S_{0}\cdots S_{k-1} S_{k}\,,\nonumber
}

\noindent with each matrix $S_k$ being a unitary transformation that has to be constructed. The strategies for constructing $S_k$ will be discussed in a moment. From the definitions in Sec. \ref{sec:Param}, $A$ can be factored as a diagonal matrix $D$ and a unitary transformation $U$, as $A=U\ D\ U^\dagger$. This relation can be inverted in order to express $D$ as a function of $A$ and $U$,

\eq{\label{eq:D}
D=U^\dagger\ A\ U
}

\noindent which is recognizable as the limit of Eq. \ref{eq:Ak} when $A=A_0$, with $A_k\rightarrow D$ and $S_0\cdots S_k\rightarrow U$. From the perspective of a numerical approximation, one may stop the sequence $\left\{A_k\right\}$ when the condition $d_k^2\le \varepsilon^2$ is met, for an arbitrary precision $\varepsilon$. In this case, Eq. \ref{eq:Ak} can be read as

\eq{
D&\approx& A_k = S^\dagger\, A_0\, S\ ,\mbox{for large enough }k,
}
\noindent and
\eq{
U&\approx& S\ \mbox{, with }\ S\equiv S_0\,S_1\,\ldots S_k\ ,
}

\noindent with a global truncation error $E\leq\varepsilon$.

Several strategies are available for constructing the sequence of rotations that satisfies these definitions. The total computational complexity depends on the number of steps in the sequence and which decisions are considered between each one. In particular, the $S_k$ can be organized in groups called {\em sweeps} where all the off-diagonal elements are systematically rotated way, one by one, in what is known as the \ac{CJM}\cite{Hansen1963,Forsythe1960,Nazareth1975}. This strategy requires no decision making regarding the elements themselves, thus employing the least amount of time per step. Another known strategy is to eliminate the largest remaining off-diagonal element with each rotation, which is Jacobi's original strategy \cite{Jacobi1846}. This strategy is proven to have quadratic convergence\cite{Golub2000}, at the cost of a search for the largest element, between rotations. This is the implementation chosen for this work, which was confirmed to achieve quadratic convergence, with the test and its results presented on the Appendix \ref{sec:Eff}. For a modern review and variations on the implementation presented here, please refer to Refs. \cite{NumRecipesCpp,Golub2000}, and references therein. 

A Jacobi rotation $S_k$ represents a single step in the process of diagonalizing the target matrix $A$ and, according to the chosen strategy, it is applied on the largest off-diagonal element, $A_{rc}$, with $r\ne c$. 
Each $S_k$ can be decomposed into two consecutive rotations, $S_k=K^{[rc]}\,G^{[rc]}$, which was first introduced by W. Givens\cite{NumRecipesCpp}, with $K$ and $G$ known as {\em Givens rotations}\cite{NumRecipesCpp}. Each of these independent rotations $K$ and $G$ is responsible for rotating away one of the two degrees of freedom of this element, since $A_{rc}$ is a complex number.  In other words, the first rotation $A'= K^{\dagger[rc]} A K^{[rc]}$ makes the resulting element $A'_{rc}$ real, while the second one $A''= G^{\dagger[rc]} A' G^{[rc]}$ is responsible for vanishing with $A''_{rc}$. Under these requirements, $K^{[rc]}$ may be written as

\begin{equation}\label{eq:Krc}
\!\!\!\!\!\! K^{[rc]}=\frac{1}{\sqrt{2}}\left[\begin{array}{ccccc}
1 &  &  &  & \\
& e^{i\theta_1} & \cdots & e^{-i\theta_1} & \\
 & \vdots & 1 & \vdots & \\
&-e^{i\theta_1}& \cdots & e^{-i\theta_1} & \\
& &&   & 1
\end{array}\right],
\end{equation}

\noindent with the main elements given by $K^{[rc]}_{rr}=e^{+i\theta_1}/\sqrt{2}=K^{\star[rc]}_{cc}$, $K^{[rc]}_{rc}=e^{i\theta_1}/\sqrt{2}=-K^{\star[rc]}_{cr}$, and $K^{[rc]}_{jj}=1$ for the diagonal elements, except at $rr$ and $cc$. All other elements are zero. By imposing that $Im\{A'_{rc}\}=0$, the rotation angle $\theta_1$ becomes

\eq{\label{eq:theta1}
\tan 2\theta_1 = \frac{Im\left\{A_{rc}\right\}}{Re\left\{A_{rc}\right\}}\ .
}

\noindent One can verify that, in the case where the target matrix $A$ is already real, $\theta_1=0$ and $K^{[rc]}$ becomes the identity $1$. The second transformation $G^{[rc]}$ must be a real rotation,

\eq{\label{eq:Grc}
G^{[rc]}=\left[\begin{array}{ccccc}
1 &  &  &  & \\
&\cos\theta_2& \cdots & \sin\theta_2 & \\
 & \vdots & 1 & \vdots & \\
&-\sin\theta_2 & \cdots & \cos\theta_2 & \\
& &&   & 1
\end{array}\right],
}

\noindent with notation analogous to the one used on Eq. \ref{eq:Krc},  $G^{[rc]}_{rr}=\cos\theta_2=K^{[rc]}_{cc}$, $K^{[rc]}_{rc}=\sin\theta_2=-K^{[rc]}_{cr}$, with $K^{[rc]}_{jj}=1$ for the diagonal elements, except at $rr$ and $cc$, and all others being zero. The rotation angle\footnote{Extra care should be taken when calculating the angles $\theta_1$ and $\theta_2$ since Eqs. \ref{eq:theta1} and \ref{eq:theta2} are prone to overflow when numerically evaluating $\tan^{-1}$.} $\theta_2$ responsible with vanishing with the element in position $rc$ is given by

\eq{\label{eq:theta2}
\tan 2\theta_2 = \frac{2\, A'_{rc}}{A'_{cc}-A'_{rr}}\ .
}

The final implementation was tested with random $3x3$ matrices, so that the numerical results could be compared to the analytical ones (see Appendix \ref{sec:Test}). In the next section, the discussion returns to how to preserve the eigenvalues ordering, making the Jacobi's algorithm suitable for studying adiabatic matter effects in neutrino physics.
\bigskip

\section{Sequential Diagonalization Strategy}\label{sec:SeqDiag}

As previously discussed in Sec.\ref{sec:Param}, once the parametrization of a Hermitian matrix is define, the objective is to obtain its eigensystem as a function of a given model's parameters, correcting for the randomness in the eigensystem ordering. The solution proposed here was loosely inspired by the \emph{parallel transport} of tangent vector, in Riemannian geometry. Since the change in both eigenvalues and eigenvectors are continuous, it should be possible to detect any unwanted reordering by simple comparison of neighboring results. This requires connecting the point of study in space parameter with another where the ordering is known, diagonalizing and correcting the ordering along the way, transporting the eigensystem from one point to the other. This method will be referred to as {\em Sequential Diagonalization Strategy} (\ac{SDS}) which can be summarized as:
\\

\begin{itemize}
    \item Starting from a point in parameter space where the eigensystem is known (including ordering), a small step is taken to a new position;
    \item The eigensystem is obtained in this new position and a comparison is drawn between the original set of eigenvalues and the new ones;
    \item Assuming that the step is short enough, it is always possible to arrive at an one-to-one match between the two sets, which allows the post-step eigensystem to be \emph{reordered} following their pre-step counterparts;
    \item This is now regarded as a new reference value and the process is repeated over a predefined path, transporting the known ordering along it.
\end{itemize}

\begin{figure}[ht]
    \centering
    \includegraphics[width=0.8\textwidth]{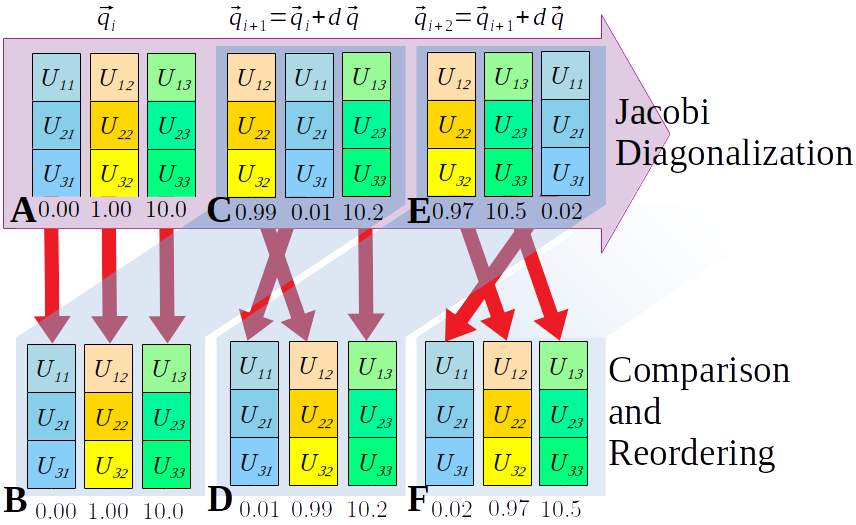}
    \caption{Graphical summary of the Sequential Diagonalization Strategy (\ac{SDS}). Each block shows the resulting diagonalizing transformation $U$ is shown, where each column represents an eigenvector, with its corresponding eigenvalue marked below it. {\bf A} is the starting point, representing the ordering to be preserved. It is stored as the first reference, shown in {\bf B}; After a small step $d\vec{q}$, a new eigensystem is obtained with the Jacobi method, shown as {\bf C}. This is compared with the previous reference, using Eq. \ref{eq:Delta}, in order to know the correct way to reorder {\bf C} to match {\bf B}; {\bf C} is stored as a new reference and the cycle repeats for the next step {\bf E}. In summary, while {\bf A}, {\bf C} and {\bf E} result from the diagonalization, with random ordering, {\bf B}, {\bf D} and {\bf F} are the values actually stored. 
    \label{fig:method}}
\end{figure}

\noindent Figure \ref{fig:method} illustrates this strategy with a graphical example. In what follows, a formal definition of \ac{SDS} is presented.

From Sec. \ref{sec:Param}, the relation $A=U\ D\ U^\dagger$ defines the eigensystem of $A$ as $\{D,U\}$, where its eigenvalues are represented as the matrix $D(\vec{q})\equiv\mbox{diag}(\lambda_1\cdots\lambda_n)$, with $\lambda_k(\vec{q})$, while its collection of eigenvectors $V_k(\vec{q})$ is organized as the columns of $U(\vec{q})$ as defined in Eq. \ref{eq.U_Definition}. In order to obtain $\{D,U\}$ at a given point $\vec{q}_\star\neq\vec{q}_0$, a path between the two points is drawn, $\vec{q}(t)$, as a function of a single parameter $t$. This new parametrization has no physical meaning and it is unrelated to how $A$ is parametrized over $\vec{q}$. This relation is introduced to reduce the number of \ac{d.o.f.} from $p$ parameters to a single one, $t$. The path is then divided into smaller steps $\delta t\ll 1$ in a total of $N=1/\delta t$. Without any loss of generality, one might consider a straight line,

\eq{\label{eq:qt}
\vec{q}(t) = \vec{q_0}+\left(\frac{t-t_i}{t_f-t_i}\right)(\vec{q_\star}-\vec{q_0})
}

\noindent as long as $\vec{q}(t)$ never leaves the domain of $A(\vec{q})$. When this is not feasible, Eq. \ref{eq:qt} can be generalized by a series of line segments or a curve of any kind. Nevertheless, the reader should keep in mind that the resulting $\{D,U\}$ are independent of the taken path, so the curve $\vec{q}(t)$ should be as simple as possible. From this point forward consider that Eq.\ref{eq:qt} is enough to define $\vec{q}(t)$. In this case, a short step $\delta t$ leads to a step $\delta \vec{q}$, with

\eq{
\delta \vec{q} = \left(\vec{q_\star}-\vec{q_0}\right)\delta t
}

\noindent with

\eq{
\vec{q}_{i+1} = \vec{q}_{i} + \delta \vec{q}
}

\noindent being the discrete representation of the chosen path. At any given step $i$, the next step will lead to $A_{i+1} =A(\vec{q}_{i+1})=A(\vec{q}_{i}+\delta \vec{q})$. When diagonalized, the resulting eigenvalue set $\{\lambda^{(i+1)}_k\}$ should have the form

\eq{
\lambda^{(i+1)}_k = \lambda^{(i)}_k + \delta \lambda_k\ .
}

\noindent This relation can be used as a tool to correct for the ordering of $\{\lambda^{(i+1)}_k\}$, by defining the quantity

\eq{\label{eq:Delta}
\Delta = \sum_{k=0}^{n}\left|\lambda^{(i+1)}_k - \lambda^{(i)}_k\right|
}

\noindent it is possible to search for the correct ordering of $\lambda^{(i+1)}$ among all possible permutations. Given that $\delta t$ is small enough, the relation $\Delta < (n\ \mbox{max}\{\lambda_k\})$ can only be true if the ordering of $\{\lambda^{(i+1)}_k\}$ matches the previous one for $\lambda^{(i)}_k$. This is performed by simple inspection, placing all permutations of $\{\lambda^{(i+1)}_k\}$  in the definition of $\Delta$ and choosing the smallest one. Once the correct permutation is known, both $\{\lambda^{(i+1)}_k\}$ and $\{V^{(i+1)}_k\}$ can be reordered and stored in $\{D_{i+1},U_{i+1}\}$. This procedure is repeated until the endpoint is reached, leading to the desired $\{D_\star,U_\star\}$.

The combination of \ac{SDS} and Jacobi's diagonalization will be referred to as \ac{SJD}. Although the \ac{SDS} can be used with any diagonalization method, it is worth noting that Jacobi's is the most suitable one for neutrino physics since it offers the possibility of evaluating both $D$ and $U$ at the same time with precision $\varepsilon$, predefined only by the stopping condition. In fact, since no other information is kept from one point to the next, besides the ordering, there are no cumulative numerical errors involved. In other words, the only errors affecting $\{D_\star,U_\star\}$ are those coming from the last diagonalization, at the point $\vec{q_\star}$ (see Appendix \ref{sec:Eff} for an in-depth discussion about precision).

A few remarks are in order. A major advantage of the \ac{SDS} is that all the diagonalizations over a path can be computed in concurrently, with the reordering done afterwards, in a serialized fashion. Also, if the intention is to map the eigensystem over a volume of parameter space, finding a way to run over such space in a continuous manner becomes a trivial task. Yet, this method is not without its limitations. The \ac{SDS} relies on the premise that there is a reference order. As a consequence, the eigenvalues have to be non-degenerate to begin with. Not only that, they also have to be different enough so Eq.\ref{eq:Delta} is applicable. It could be the case, however, that some particular parametrization causes two or more eigenvalues to cross each other, becoming degenerate at that point. This kind of ambiguity can be solved by adopting a higher order discriminant, such as comparing $\lambda^{i+1}_k$ with $\lambda^{i-1}_k$, which is equivalent to comparing the derivatives of $d\lambda_k/dt$. 
\bigskip

\section{Neutrino Physics Application}\label{sec:Neutrinos}

This section offers an example application of \ac{SJD} in neutrino physics. The goal is to obtain the mixing (oscillation) parameters, defined by the \ac{PMNS} parametrization, as a function of the matter background in the \ac{MSW} effect. (see Refs. \cite{Smirnov2016,Bilenky2014} for a modern review). In it, the presence of an interacting medium shifts the energy levels of the Hamiltonian, which in turn leads to a new set of effective values for the neutrino mixing parameters, either enhancing or suppressing the oscillation pattern, depending on the matter profile along the propagation path. In what follows, the \ac{MSW} effect is briefly reviewed and the usage of the \ac{SJD} is illustrated for a 3-neutrino, \ac{SM} case.
\\

\begin{table}[hb]
\centering
\begin{tabular}{llccc}
\toprule
\textbf{Parameter} & \textbf{Best fit} & \textbf{$1\sigma$ range} & \textbf{$3\sigma$ range} \\
\midrule
\vspace{-0.2cm}\\
$\Delta m^2_{21}$ {\footnotesize $\!/10^{-5}$} & 7.37 & 7.21\,-\,7.54 & 6.93\,-\,7.97\\
\vspace{-0.1cm}\\
$\sin^2 \theta_{12}$ {\footnotesize $\!/10^{-1}$}& 2.97 	&	2.81\,-\,3.14	 	& 2.50\,-\,3.54 	\\
\midrule
\multicolumn{4}{c}{\textbf{ \ac{NH} }}\\
\midrule
\vspace{-1em}\\
$+\Delta m^2_{31}$ {\footnotesize $\!/10^{-3}$}	&	2.39	& 2.35\,-\,2.43 	& 2.27\,-\,2.51 	\\
\vspace{-0.1cm}\\
$\sin^2 \theta_{13}$ {\footnotesize $\!/10^{-2}$}&	2.14	& 2.05\,-\,2.25	 	& 1.85\,-\,2.46 	\\
\vspace{-0.1cm}\\
$\sin^2 \theta_{23}$ {\footnotesize $\!/10^{-1}$} & 4.37 & 4.17\,-\,4.70 	& 3.79\,-\,6.16 	\\
\vspace{-0.1cm}\\
$\delta_{CP}/\pi$ & 1.35 & 1.13\,-\,1.64 & 0\,-\,2 \\
\midrule
\multicolumn{4}{c}{\textbf{ \ac{IH} } }\\
\midrule
\vspace{-1em}\\
$-\Delta m^2_{31}$ {\footnotesize $\!/10^{-3}$}	&	2.35	& 2.31\,-\,2.40 	& 2.23\,-\,2.48 	\\
\vspace{-0.1cm}\\
$\sin^2 \theta_{13}$ {\footnotesize $\!/10^{-2}$}&	2.18	&	2.06\,-\,2.27	 	& 1.86\,-\,2.48 	\\
\vspace{-0.1cm}\\
$\sin^2 \theta_{23}$ {\footnotesize $\!/10^{-1}$} & 5.69 & 4.28\,-\,4.91 	& 3.83\,-\,6.37 	\\
\vspace{-0.1cm}\\
$\delta_{CP}/\pi$ & 1.32 & 1.07\,-\,1.67 & 0\,-\,2 \\
\bottomrule
\end{tabular}
\caption{Best fit values, $1\sigma$ and $3\sigma$ rages for the global fit of all relevant neutrino oscillation data\cite{CAPOZZI2016}. Here, the notation used by the original reference is changed in favor of one that best suits this work.}\label{tab:SM}
 \end{table}

As the mass-flavor mixing model states\cite{Bilenky2014}, a three-neutrino system can be represented by a free Hamiltonian which is diagonal when expressed in the mass basis,  $H_m=(\Delta m^2_{21}/2p)\times\mbox{diag}\left(0,1,\alpha \right)$, with $\alpha=\Delta m^2_{31}/\Delta m^2_{21}$ and the $\Delta m^2_{ij}$ are the squared-mass differences between the neutrino mass-states. The unitary mixing matrix $U$ takes the diagonal $H_m$ to the flavor basis via a similarity transformation,

\eq{
    H_f=U\ H_m\ U^\dagger\ ,
}

\noindent with $U$ being the result of three real rotations (with Euler angles $\theta_{12}$, $\theta_{23}$ and $\theta_{13}$) and at least one complex phase $\dCP$. In the presence of an interacting background, represented by a potential matrix $V$, the total Hamiltonian of the system becomes

\eq{
\tilde{H}_f=U\ H_m\ U^\dagger + V\ ,
}

\noindent where the $\sim$ sign represents non-vacuum values, with $V$ being a general real matrix, encoding how each neutrino flavor interacts with the medium. The physical observables are those related to the neutrino oscillation pattern, namely the oscillation length and amplitudes, given by the eigenvalues and eigenvectors of $\tilde{H}_f$, respectively. Let $\tilde{U}$ be the diagonalizing transformation that realizes the following,

\eq{\label{eq:Hmtilde}
\tilde{H}_m &=& \tilde{U}^\dagger\ \tilde{H}_f\ \tilde{U} \\
&=& \tilde{U}^\dagger\ \left( U\ H_m\ U^\dagger + V \right) \tilde{U}\nonumber
}

\noindent where $\tilde{H}_m\rightarrow H_m$, and $\tilde{U}\rightarrow U$, when $V\rightarrow 0$. Eq. \ref{eq:Hmtilde} is equivalent to Eq.\ref{eq:D}, with $A=\tilde{H}_f$ and $D=\tilde{H}_m$. This single realization evokes the motivation behind this study, since while Eq.\ref{eq:D} is just the starting point of a diagonalization tool, Eq. \ref{eq:Hmtilde} has actual meaning in neutrino physics.

Consider $\tilde{H}_m=(\Delta m^2_{21}/2p)\times\mbox{diag}\left(\lambda_1,\lambda_2,\lambda_3 \right)$ as the diagonal form of $\tilde{H}_f$, with $\lambda_k$ the relevant factors of its eigenvalues. From the elements of the diagonalizing transformation of $U$ (or $\tilde{U}$) it is possible to define three Mixing Amplitudes, 

\eq{
    \sin^2 2\theta_{12} = 4\frac{\left|U_{e1}\right|^2\,\left|U_{e2}\right|^2}{\left(1-\left|U_{e3}\right|^2\right)^2}\ ,\mbox{\hspace{1cm}}\ 
    \sin^2 2\theta_{23} =4\frac{\left|U_{\mu 1}\right|^2\,\left|U_{\tau 2}\right|^2}{\left(1-\left|U_{e3}\right|^2\right)^2}\ ,    
    \nonumber
}

\noindent and
\eq{\label{eq:amplitudes}
\sin^2 2\theta_{13} &=& 4\left|U_{e3}\right|^2\,\left(\left|U_{\mu 3}\right|^2+\left|U_{\tau 3}\right|^2\right)\, .
}

\noindent representing how the mass-eigenstates are mixed into the flavor states. This notation is corresponds to the \ac{PMNS} parametrization\cite{Smirnov2016}. It is also possible to isolate the effects of CP-violation, represented by the Jarlskog invariant, 

\eq{\label{eq:JCP}
\JCP=Im\left\{U_{\mu3}\,U^\star_{\mu2}\,U_{e2}\,U^\star_{e3}\right\}
}

\noindent where $\JCP$ represents the difference between neutrino and anti-neutrino oscillation. 

\begin{figure}[htbp]
    \centering
    \begin{subfigure}{0.48\textwidth}
        \includegraphics[width=\textwidth]{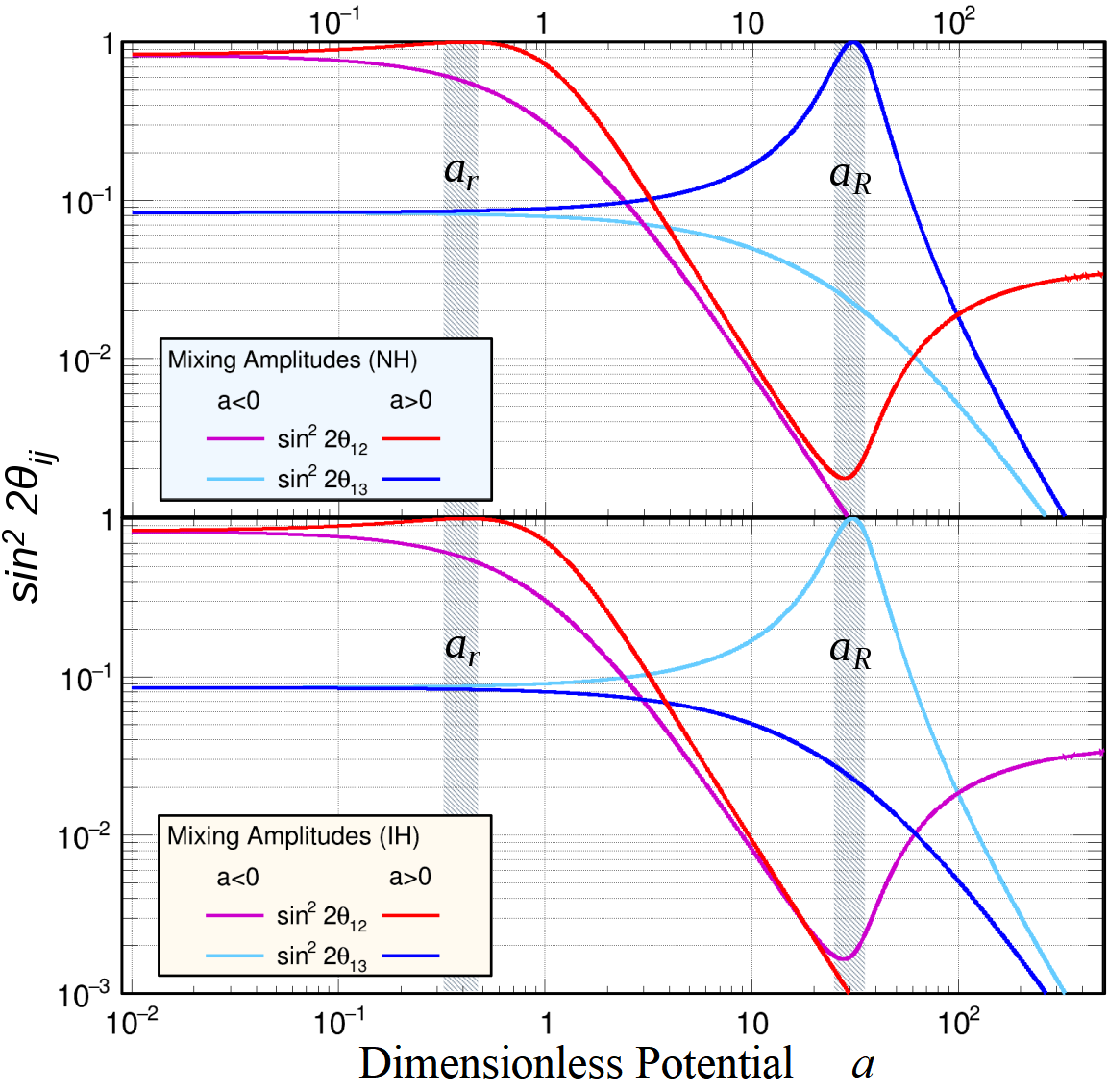}    
        \caption{Mixing Amplitudes, as a function of $|a|$.} \label{fig:amplitudes_12_13}
    \end{subfigure}
    \hfill
    \begin{subfigure}{0.46\textwidth}
        \includegraphics[width=\textwidth]{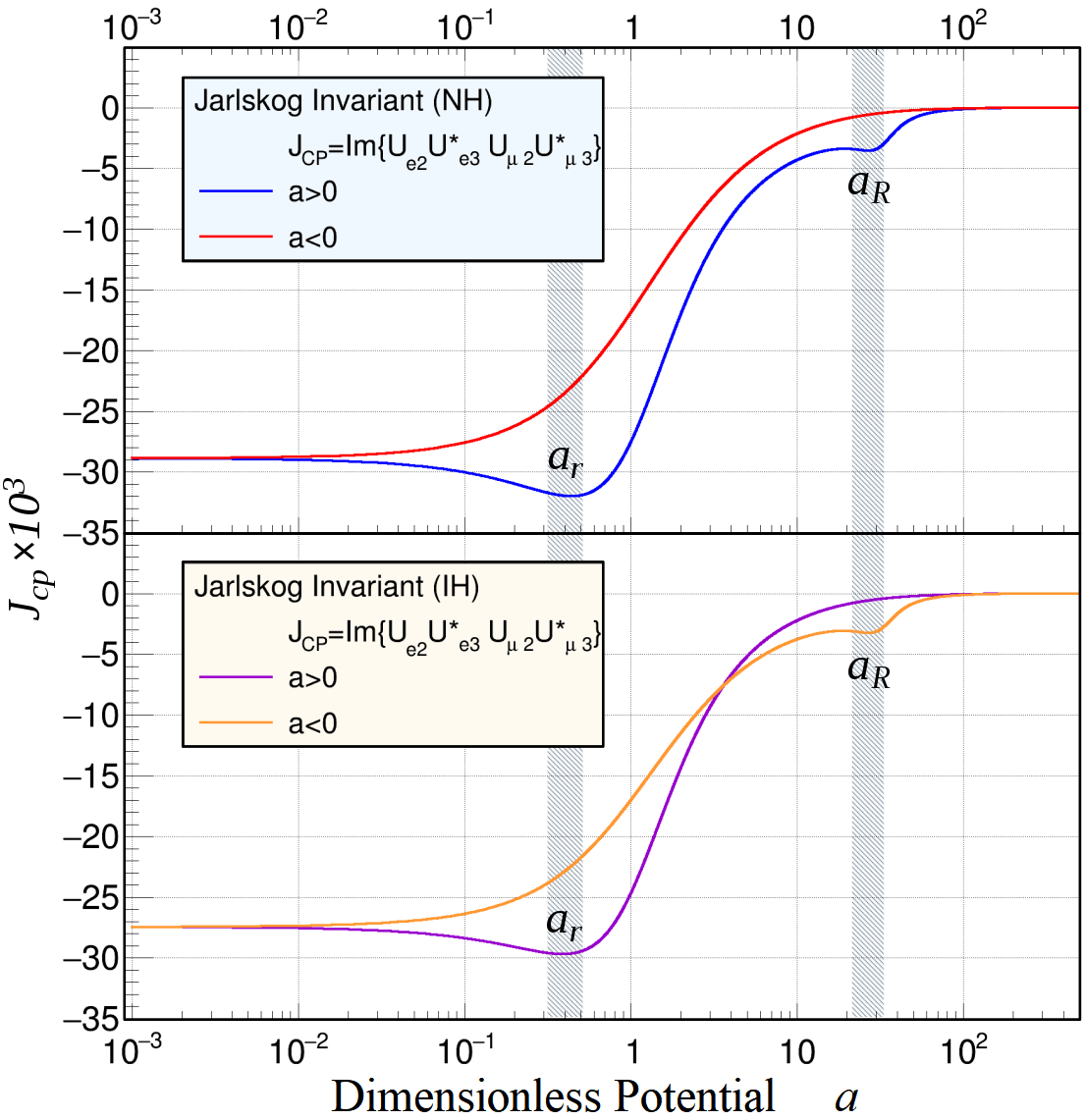}    
        \caption{Jarlskog Invariant, as a function of $|a|$.} 
        \label{fig:Diagonal_Jcp}
    \end{subfigure}
    \\
    \vspace{0.5cm}
    \caption{(a) The matter-enhanced values of $\sin^2 2\theta_{12}$ and $\sin^2 2\theta_{13}$, with red and blue for $a>0$ and magenta and cyan for $a<0$, respectively, with \ac{NH} at the top plot and \ac{IH} at the bottom. (b) The matter enhanced values of $\Jcp$, as defined on Eq.\ref{eq:JCP}.  At the top is the \ac{NH} scenario, with blue for $a>0$ and red for $a<0$, while at the bottom is the \ac{IH} case, with magenta for $a>0$ and orange for $a<0$.}
\end{figure}

The angles in Eqs. \ref{eq:amplitudes} and the Jar \ref{eq:JCP} will lead to different values, depending on the elements of the potential $V$. As an example, in non-standard interaction searches, the elements of $V$ can be either independent of each other or given by an underling model. In the case of an ordinary-matter background, however, the potential can be as simple as $V=(\Delta m^2_{21})/2p\times\mbox{diag}\left(a,0,0 \right)$, with $a=2p V_{cc}/\Delta m^2_{21}$ (more on  $V_{cc}$ in a moment). Regardless of the model, being it \ac{SM} or \ac{BSM}, \ac{SJD} can be used to obtain the behavior of $\tilde{H}_f$'s eigensystem as a function of a specific model parameter or even the complete set of $V$ elements, which would be model-independent. In the case of a constant and uniform background, these definitions are enough to completely define the system. When this is not the case, it becomes necessary to also know how  $\tilde{U}$ varies along the neutrino's trajectory $x$, i.e.,

\eq{\label{eq:dU}
\frac{d\tilde{U}_{ij}}{dx} = \sum_{k\ell}\ \frac{dU_{ij}}{dV_{k\ell}}\ \frac{dV_{k\ell}}{dx}\ .
}

\noindent This is the scenario where the \ac{SJD} can provide sizable improvement over other methods. The values of ${d\tilde{U}_{ij}}/dx$ can be obtained prior to a full model analysis since they should be recalculated less frequently, if ever twice in a single study. Monte Carlo productions, as well as model fitting, can make use of a \ac{LUT} instead of performing thousands of diagonalizations at every step. The more demanding simulations for the next generation of neutrino detectors, such as \ac{DUNE}\cite{DUNE2020}, should benefit from this approach.

To take a concrete example, we can appreciate an application using only Standard Model physics, for which there are analytical solutions. In this case, the relevant matter potential is $V=\mbox{diag}\left(V_{cc},0,0\right)$, with $V_{cc}$ being the charged current potential between electrons in the medium and the electron-(anti)neutrino. For neutral baryonic matter, $V_{cc}=\sqrt{2}G_F\, n_e$, where $n_e$ is the background's electron density and $G_F$ is Fermi's constant. Since global phases do not influence the final oscillation probabilities, we can place $\Delta m^2_{21}/2p$ in evidence, writing $V=\mbox{diag}\left(a,0,0 \right)$, with $a=2p V_{cc}/\Delta m^2_{21}$. The parameter $a$ encodes all the background description such as density, interaction strength, and uniformity. By using \ac{SJD}, it is possible to obtain all the relevant observables and still be agnostic with respect to the background properties, which can be added at a later point of the computation. 

The total Hamiltonian to be diagonalized is $\tilde{H}_m(a)$, where $a>0$ means that both the neutrinos and the background are of the same nature, i.e., either both matter or both antimatter, while $a<0$ represents the matter/antimatter combination. Using the vacuum values on \ref{tab:SM}, it is possible to obtain two distinct values for $\alpha$, $\alpha_{\mbox{\scriptsize NH}}=32.4$ and $\alpha_{{\mbox{\scriptsize IH}}}=-31.9$, corresponding to Normal Hierarchy (\ac{NH}) and Inverted Hierarchy (\ac{IH}), respectively. All results that follow will show four distinct cases: $\pm a$ and \ac{NH}/\ac{IH}.

The analytical solutions for the Mixing Amplitudes\cite{Zaglauer1988,Kimura2002} are compared to the \ac{SJD}, being in agreement up to the chosen precision ($\varepsilon=10^{-14}$). Fig. \ref{fig:amplitudes_12_13} shows the mixing amplitudes $\sin^2 2\theta_{12}$ and $\sin^2 2\theta_{13}$ as a function of $|a|$ ($\sin^2 2\theta_{23}$ is not shown since it is indistinguishable from 1 in this scale). It is possible to observe the resonant \ac{MSW} effect, related to the two mass-scales. The lower resonance $a_r$ affects $\sin^2 2\theta_{12}$, while the higher one $a_R$ affects $\sin^2 2\theta_{13}$ and $\sin^2 2\theta_{23}$. The latter is not shown on the plots since it would be indistinguishable from unit due to its large vacuum value. 

\begin{figure}[ht]
  \centering
  \begin{subfigure}{0.45\textwidth}
    \centering
    \includegraphics[width=\textwidth]{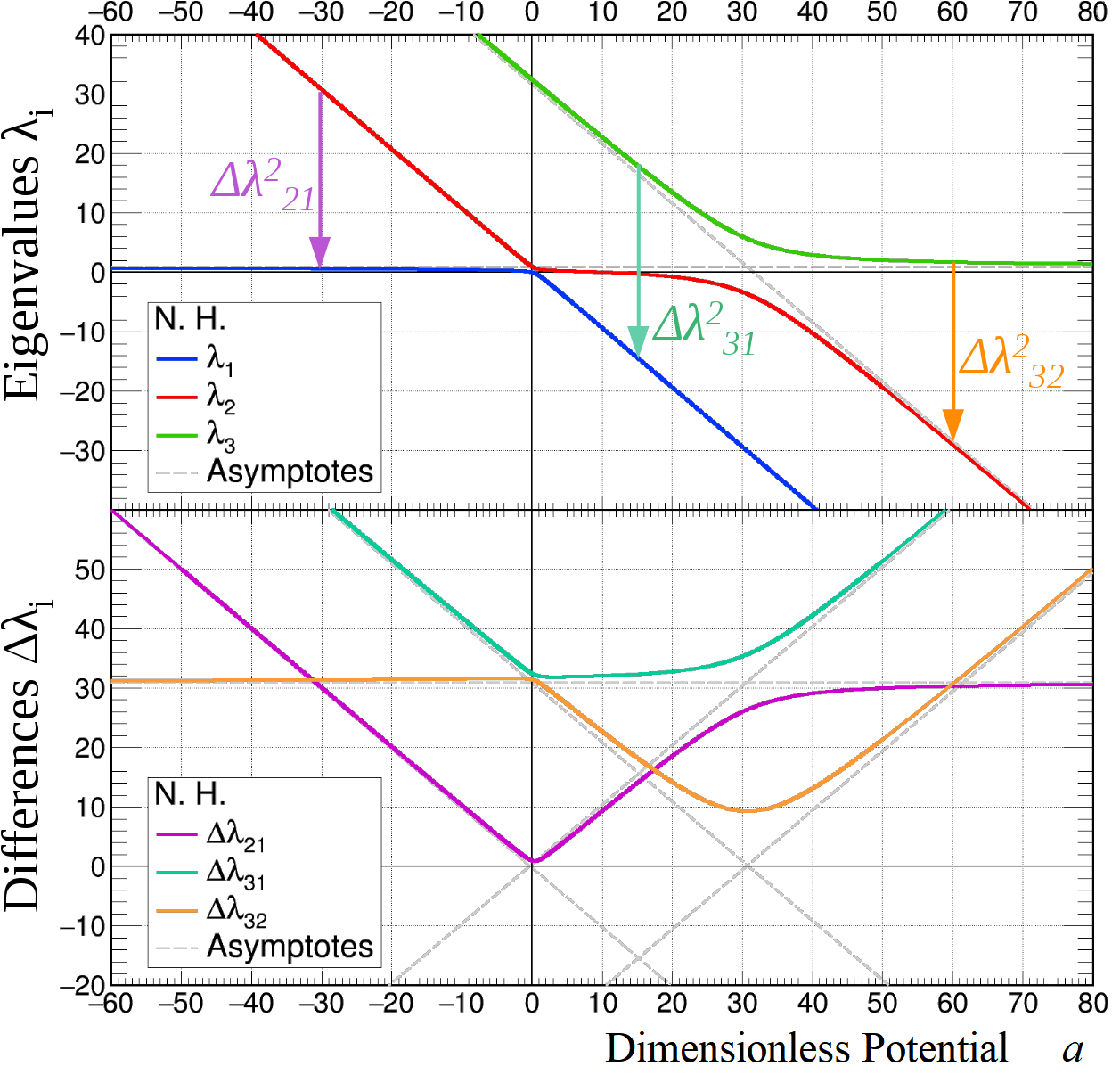}
    \caption{Eigenvalues of $\tilde{H}_m$ for \ac{NH}.}
    \label{fig:eigenvaluesNH}
  \end{subfigure}
  \hfill
  \begin{subfigure}{0.45\textwidth}
    \centering
    \includegraphics[width=\textwidth]{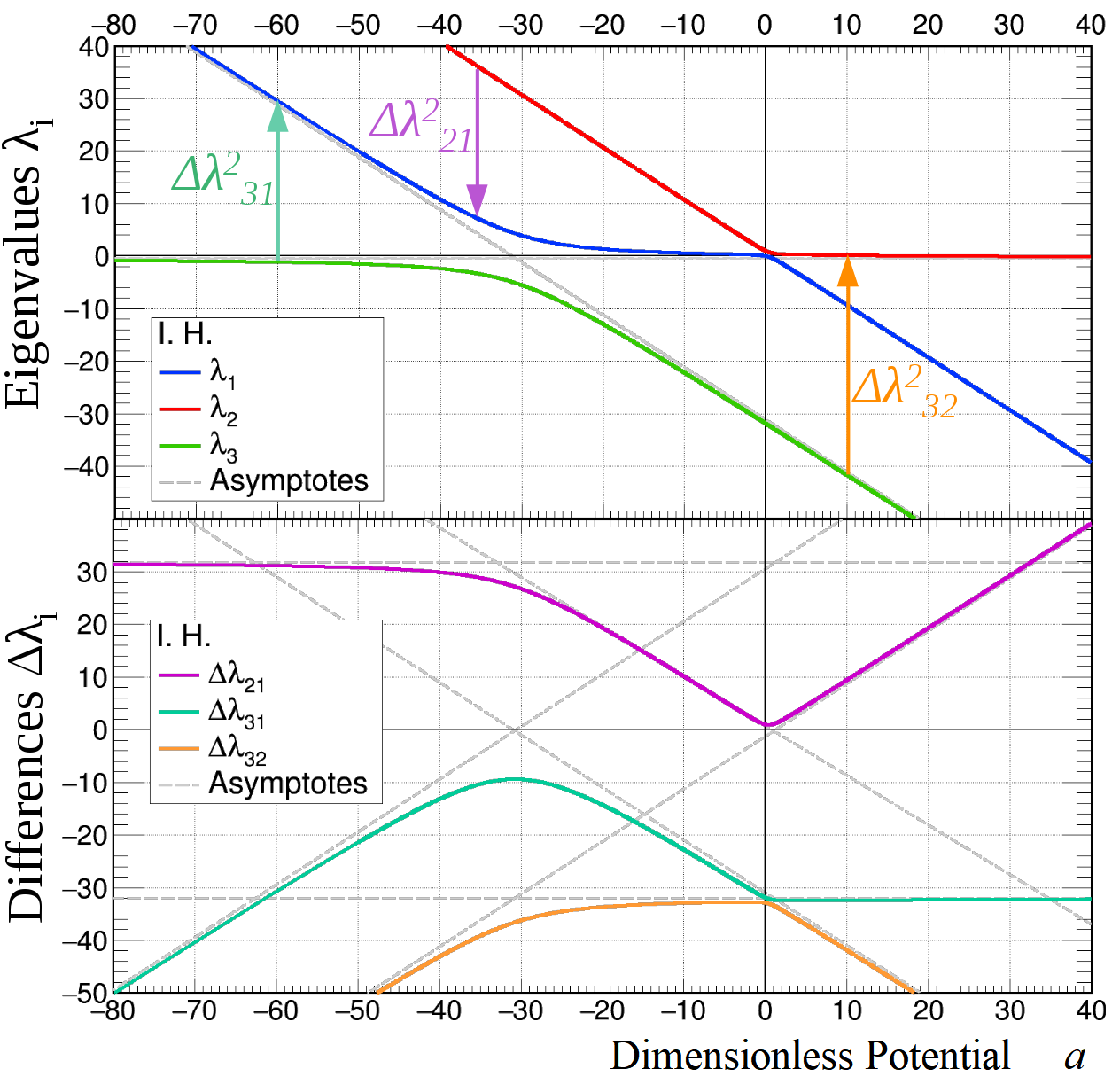}
    \caption{Eigenvalues of $\tilde{H}_m$ for \ac{IH}.}
    \label{fig:eigenvaluesIH}
  \end{subfigure}
  \bigskip
  
    \caption{{Eigenvalues of $\tilde{H}_m$ (Eq.\ref{eq:Hmtilde}), for \ac{NH} (a) and \ac{IH} (b)}. Each individual eigenvalue $\lambda_i$ is drawn as a function of the dimensionless background descriptor $a$, representing the \ac{MSW} effect in ordinary matter. The differences $\Delta\lambda_{ij}$ are also shown, since they correlate to the oscillation length. All vacuum values are taken from (Tab \ref{tab:SM}).} \label{fig:eigenvalues}
\end{figure}

In Fig. \ref{fig:eigenvalues}, we observe the eigenvalues of $H_m$ for \ac{NH} and \ac{IH}. The vacuum eigenvalues are $\lambda_1=0$, $\lambda_2=1$, and $\lambda_3=\alpha$, and it is possible to see that resonances $a_r$ and $a_R$ represent the points where the eigenvalues change asymptotes.

Finally, Fig. \ref{fig:Diagonal_Jcp} shows how the Jarskog invariant is affected by the background. Regardless of the hierarchy case, $a_r$ represents a resonant minimum for $a>0$, and $a>a_R$ will always lead to $\JCP=0$, meaning that neutrinos and antineutrinos would behave the same. It is worth noting that the existence of \ac{CP}-violation in neutrino oscillations is not confirmed and the results shown here only consider the best fit values for $\dCP$, which is still compatible with zero. No matter what the true $\dCP$ is, it affects $\JCP$ with all the previous observables remaining unchanged.

\section{Conclusions}\label{sec:Conc}

The Sequential Jacobi Diagonalization, or \ac{SJD} proposed in this work combines a heuristic procedure with a well established numerical method in order to satisfy the computation requirements for neutrino physics application. In this field, computational resources become a bottleneck whenever \ac{BSM} hypotheses are being tested. In more general terms, given the description of a Hermitian system, modeled over a particular set of parameters, this method allows for the study of how the eigenvalues and eigenvectors are related to these parameters.

\section*{Acknowledgments}

The author wishes to thank Prof. Dr. Alberto M. Gago (ORCID 0000-0002-0019-9692) for his insightful and encouraging comments on this work. Also to acknowledge funding retrieved from {\em Conselho Nacional de Desenvolvimento Cient\' \i fico e Tecnol\'ogico}, CNPq (grant No. 477588/2013-1) and {\em Funca\c c\~ao de Amparo a Pesquisa do Estado de Minas Gerais}, FAPEMIG (grants No. APQ-01439-14, APQ-00544-23 and APQ-01249-24). This work was partially developed during a fellowship period granted by {\em Coordena\c c\~ao de Aperfei\c coamento de Pessoal de N\'\i vel Superior} - CAPES (grant No. 88881.121149/2016-01).




\appendix

\section{Precision and Efficiency}\label{sec:Eff}

\begin{figure}[ht]
\begin{center}
	\includegraphics[width=0.7\textwidth]{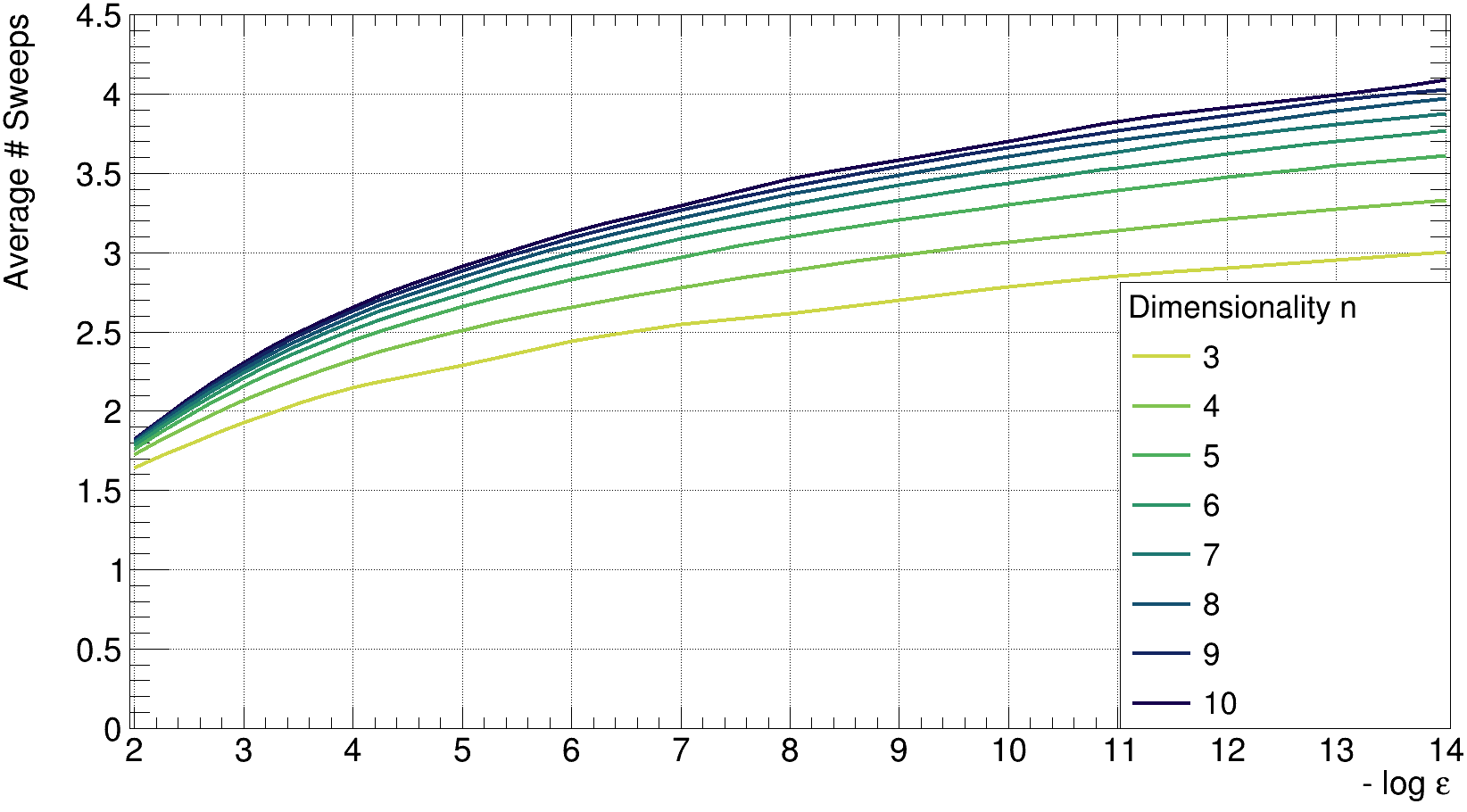}
\end{center}
\caption{Average number of sweeps before converging to a diagonal, with given precision $\varepsilon$ (shown as $-\log \varepsilon$). One sweep is defined as the application of one full complex rotation for each off-diagonal element, $n(n-1)/2$. Each line represent a dimensionality from $n=3$ to
$10$. This average is obtained from a random sample of $10^6$ Hermitian matrices with real and imaginary parts limited to the interval $[-1,+1]$. It is possible to obtain diagonalizations with the largest off-diagonal element bounded as $<10^{-14}$ by employing an average of 3 to 4 sweeps, for matrices up to $10\times10$. }\label{fig:Convergence}
\end{figure}

In order to evaluate stability, convergence and precision, I applied the Jacobi diagonalization to a sample of $10^6$ random Hermitian matrices, with the real and imaginary parts of each element constrained to $[-1,+1]$. This sample is representative since any matrix can be normalized by its largest element in order to fit in this range. After reaching the stop condition $d^2\leq \varepsilon^2$, as defined on Eq. \ref{eq:d2}, the resulting diagonal form is rotated back with the obtained $U=S^\dagger$, and compared component-wise with the original matrix. The largest difference is found to be always smaller than the target precision $\varepsilon$, meaning that $-\log_{10}\varepsilon$ is a good indicator of the number of significant figures achieved in the solutions. For each sample, $\varepsilon$ is varied from $10^{-2}$ to $10^{-16}$, where the average number of rotations $S_k$ is recorded. This virtual experiment is repeated from n = 3 to $n=10$ matrices and the results are shown in Fig. \ref{fig:Convergence}. 

Since the method targets the largest elements, not all in sequence (contrary to its cyclic variant), the average ``sweep" is defined as the ratio between the number of complex rotations (two real rotations from Eq.\ref{eq:Krc} and \ref{eq:Grc}) and the number of off-diagonal elements $n(n-1)/2$. This ratio is strictly larger than $1$, regardless of the dimensionality, since a general matrix requires at least one complex rotation for each off-diagonal element. The actual number of rotations goes with $\mathcal{O}(n^2)$.  By establishing the $d^2\leq\varepsilon^2$ limit from Eq. \ref{eq:d2} as stopping criterion, there is a possibility that some elements might underflow if the required $\varepsilon$ is too close to machine precision. Indeed, this is observed when requiring $\varepsilon\leq 10^{-15}$, using 64-bit floating-point variables (which can represent a maximum of $15$ significant figures). Stability and convergence are observed with $\varepsilon\geq 10^{-14}$, which is the largest precision shown in Fig. \ref{fig:Convergence}. In this limit, numerical diagonalization is achieved with an average of between $3$ and $4$ sweeps. This average held even for matrices as large as $10\times 10$. Most physical applications would realistic require far less precision than the $10^{-14}$ tested, which translates to a less demanding process. Tab.\ref{tab:Precision} shows the average number of sweeps, the standard deviation, and how many sweeps are needed to diagonalize $99\%$ of each sample. Even in the most demanding case, with $n=30$, a five significant figures precision can be obtained with a maximum of $3.2$ sweeps. Also, the standard deviation around this average gets narrower as $n$ increases. Both Fig.\ref{fig:Convergence} and Tab.\ref{tab:Precision} show evidence of a possible limit, or at least a log-like growth, in the number of sweeps as a function of $n$. This cannot be verified by employing only numerical analysis, so no further statements will be made on this observation. It can be said however, that the expected number of real rotations is $3n^2$, as a thumb-rule. As a final remark, quadratic convergence (precision = $\mbox{sweeps}^2$) was observed for all tested dimensionalities, as suggested by the literature\cite{Hansen1963}.

\begin{table}[ht]
\centering
\begin{tabular}{rccc}
\toprule
\textbf{n} & \textbf{Avg. Sweeps} & \textbf{Std. dev.} & \textbf{$99\%$ less than} \\
\midrule
\vspace{-0.2cm}\\
3 & 2.30 & 0.19 & 2.7 \\
4 & 2.51 & 0.19 & 3.0 \\
5 & 2.66 & 0.13 & 3.1 \\
6 & 2.74 & 0.10 & 3.1 \\
7 & 2.81 & 0.10 & 3.1 \\
8 & 2.85 & 0.09 & 3.2 \\
9 & 2.88 & 0.08 & 3.2 \\
10 & 2.92 & 0.08 & 3.2 \\
\midrule
20 & 3.07 & 0.05 & 3.2 \\
30 & 3.15 & 0.04 & 3.3 \\
\toprule
\end{tabular}
\caption{Average number of sweeps required for convergence a diagonal, with precision $\varepsilon=10^{-5}$, from the a $10^6$ random matrices sample. The first column (Avg. Sweeps) corresponds to a cut from Fig.\ref{fig:Convergence} at $-\log\varepsilon=5$. The second one (Std. dev.) shows the standard deviation from each sample. The rightmost column shows how many sweeps were needed for $99\%$ of each sample to reach the stopping condition. The two bottom rows show extra information not present in Fig.\ref{fig:Convergence}, for $n=20$ and $n=30$. Even with such large matrices, less than $1\%$ of the matrices reacquired more than $3.3$ sweeps.}\label{tab:Precision}
\end{table}

\section{Numerical vs. Analytical}\label{sec:Test}
\renewcommand{\theequation}{B.\arabic{equation}}
\setcounter{equation}{0} 

\renewcommand{\thetable}{B.\arabic{table}}
\setcounter{table}{0} 

In this section, a random example with an analytic solution is analyzed. The goal is to validate the numerical methods proposed in this work. A particular $n=3$ case with a known analytic solution is used to exemplify the validity of the method. Starting with two Hermitian matrices, $A$ and $B$, given by

\eq{\label{eq:A}
A=\left[\begin{array}{ccc}
\!\! 3 & \!\! i & \!\! 0  \\
-i & -2 & \!\! i  \\
0 & -i & \!\! 1  
\end{array}\right],
}
\noindent and
\eq{\label{eq:B}
B=\left[\begin{array}{ccc}
1 & 0 & 0  \\
0 & 2 & 0  \\
0 & 0 & 3  
\end{array}\right],
}

\noindent a linear parametrization is defined as

\eq{\label{eq:Hx}
H(x) = A\ x+B\ .
}

\noindent These were chosen among several tests for no particular reason other than to provide a good example. The eigenvalues of $B$ are not only explicit, since $B$ is diagonal by definition, but also their ordering is well determined. The eigenvalues of $A$ are obtained by solving its order-3 characteristic polynomial, leading to

\eq{
\lambda_{1}^{A} &=& \frac{1}{3}
\left[ 2-\mbox{Re}\left\{\sigma\ (1-i\sqrt{3})\right\} \right]\ , \label{eq:Lambda1A}\\
\lambda_{2}^{A} &=& \frac{1}{3}
\left[ 2+\mbox{Re}\{\sigma\} \right]\ , \label{eq:Lambda2A}
}
\noindent and
\eq{
\lambda_{3}^{A} &=& \frac{1}{3}
\left[ 2-\mbox{Re}\left\{\sigma\ (1+i\sqrt{3})\right\} \right]\ ,  \label{eq:Lambda3A}
}

\noindent where $\sigma = (-64+3i\sqrt{1281})^{1/3}$ (the first complex root). Their numbering is reflecting their relative positioning on the number line, $\lambda_{1}^{A}\geq \lambda_{2}^{A}\geq \lambda_{3}^{A}$, not parametrization ordering. For the sake of this analysis, all numerical values are quoted with $10^{-5}$ precision even when using analytical formulas. The eigenvalues of $A$ are $\lambda_{1}^{A}= -2.47090$, $\lambda_{2}^{A}= 1.26071$, and $\lambda_{3}^{A}= 3.21018$, as defined by Eqs. \ref{eq:Lambda1A}, \ref{eq:Lambda2A}, and \ref{eq:Lambda3A}.

\begin{figure}[htbp]
  \centering
  \begin{subfigure}{0.45\textwidth}
    \centering
    \includegraphics[width=\textwidth]{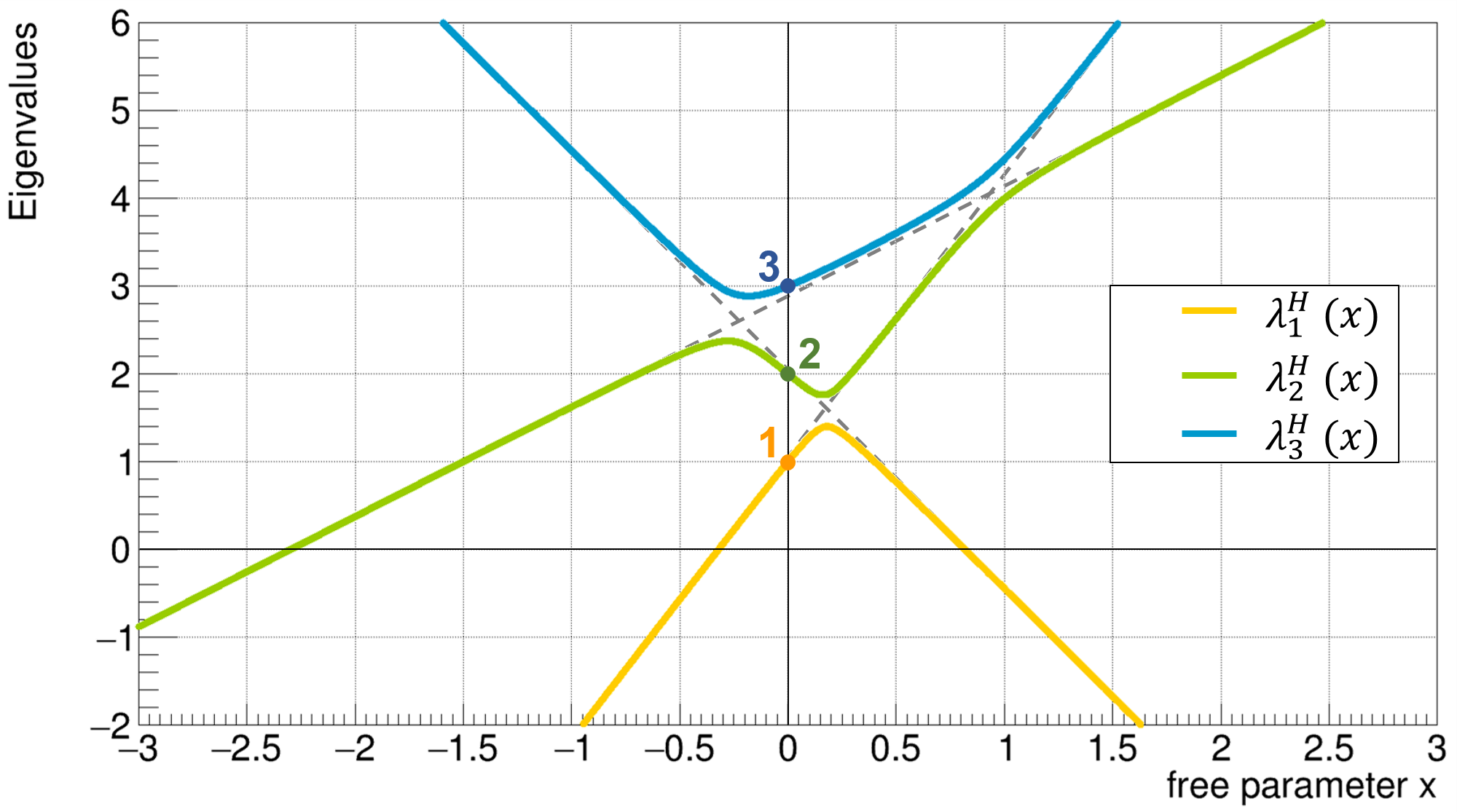}
    \caption{eigenvalues of $H(x)$.  }
    \label{fig:exEigenvalues}
  \end{subfigure}
  \hfill
  \begin{subfigure}{0.45\textwidth}
    \centering
    \includegraphics[width=\textwidth]{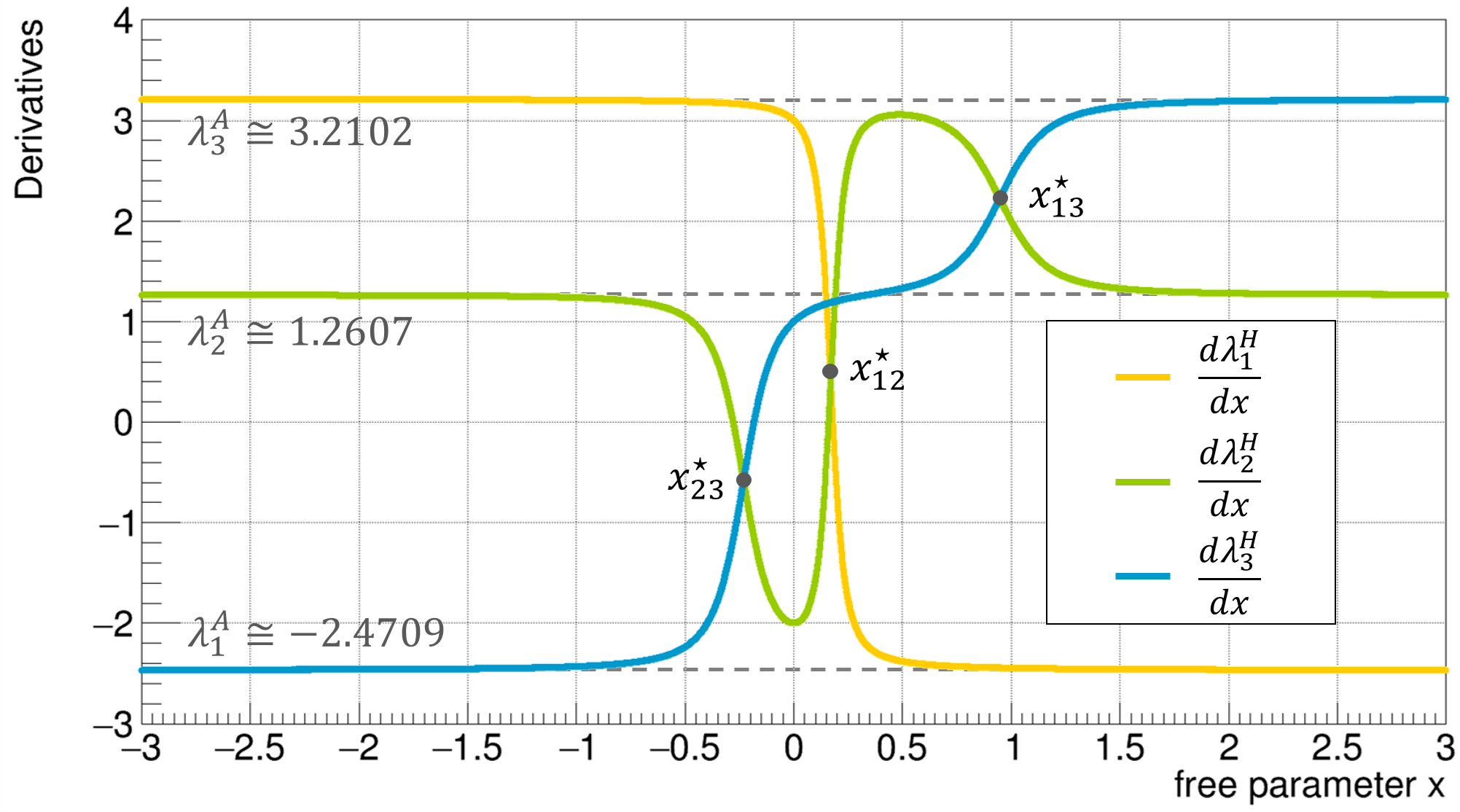}
    \caption{Derivative of the eigenvalues of $H(x)$.}
    \label{fig:exDerivatives}
  \end{subfigure}
  \bigskip
  
  \caption{(a) Eigenvalues of $H(x)$, as defined by Eqs. \ref{eq:A}, \ref{eq:B} and \ref{eq:Hx}.  The continuous lines shows $\lambda_1^H$ (yellow), $\lambda_2^H$ (green) and $\lambda_3^H$ (blue), as a function of the free parameter $x$. The dashed lines indicate the asymptotes (Eq.\ref{eq:Asy}) and their intersections (Tab.\ref{tab:Intersections}). (b)  Derivative of the Eigenvalues of $H(x)$ with respect to $x$, as defined by Eqs. \ref{eq:A}, \ref{eq:B} and \ref{eq:Hx}. The continuous lines shows $d\lambda_1^H/dx$ (yellow), $d\lambda_2^H/dx$ (green) and $d\lambda_3^H/dx$ (blue), as a function of the free parameter $x$. The dashed lines indicate the asymptotes (Eq.\ref{eq:Asy}), which correspond to the eigenvalues of $A$ listed in Eqs.\ref{eq:Lambda1A}, \ref{eq:Lambda2A}, and \ref{eq:Lambda3A}. Their intersections correspond to those of the asymptotes defined by Eq.\ref{eq:Asy} and their numerical values displayed on Tab. \ref{tab:Intersections}. }
  \label{fig:mainfig}
\end{figure}

\begin{figure}[htbp]
    \centering
    \includegraphics[width=0.5\textwidth]{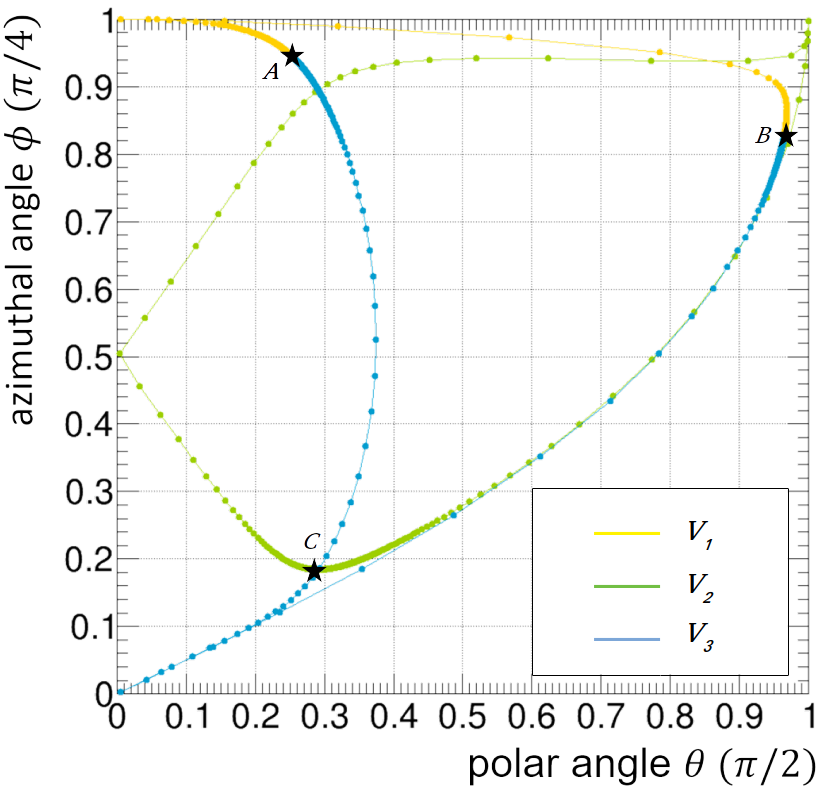}
    \caption{{Spherical representation of $H(x)$'s eigenvectors}. The eigenvectors $V_1$ (yellow), $V_2$ (green), and $V_3$ (blue). The starting point for each vector is $V_1=(0,1)$, $V_2=(1,1)$, and $V_3=(0,0)$. The dots represent equal steps in $x$, and help demonstrate the asymptotic behavior, when the density of points increase (meaning $x\rightarrow\infty$ (denoted by the three stars). The point $A=(0.25721,0.94092)$,  corresponds to either $V_1$ when $x\rightarrow -\infty$ or $V_3$ when $x\rightarrow +\infty$. Similarly, $B=(0.96496,0.82418)$ is limit of $V_1$ when $x\rightarrow +\infty$ or $V_3$ when $x\rightarrow -\infty$. And finally, $C=(0.31152,0.18597)$ is the convergence of $V_2$ for both $x\rightarrow \pm\infty$. } \label{fig:exPolar}
\end{figure}

One wishes to study the parametrized eigensystem of $H$, represented by $\{D,U\}$, as a function of $x$. By the definition in Eq.\ref{eq:Hx}, $D(x=0)=B$ therefore $U(x=0)=1$. In other words, at $x=0$ the eigenvalues of $H$ are not only the same as those of $B$, but they follow the same order. It is also possible to infer just from Eq. \ref{eq:Hx} the behavior of $H$ when $x\rightarrow \pm \infty$, since $A\, x$ becomes the dominant term and the eigenvalues of $H$ assume the form of $x\ \lambda_k^{A}$. This means that the eigensystem of $H$ has an asymptotic behavior and, for instance, one might be tempted to write $\lambda_{1}^{H}(x) =  x\ \lambda_{1}^{A} + \lambda_{1}^{B}$ in order to describe $\lambda_{1}^{H}(x)$ asymptote. However, there is no explicit information stating which $x\ \lambda_k^{A}$ corresponds to which $\lambda_j^{B}$. Unless $H$ is diagonalized, the true correspondence between the eigenvalues near zero and its value elsewhere is not clear yet, being implicitly determined by the parametrization. Besides, the same $\lambda^{H}$ can have different asymptotes for each limit.

By applying the Sequential Jacobi Diagonalization, described in Sec. \ref{sec:Jacobi} and \ref{sec:SeqDiag}, numerical representation of $D(x)$ and $U(x)$ can be calculated for a range of $x$ around the origin. Figure \ref{fig:exEigenvalues} shows the functions $\lambda_j^{H}(x)$ for a $|x|\geq 3$, which contain all of this system's features. Their behavior is analogous to that of trains changing tracks. There are three asymptotes, of the form 

\eq{
f_k & = & x\ \lambda_k^{A} + a_k ,\ \ \mbox{k=1,2,3}\ ,\label{eq:Asy}
}

\noindent where $k$ indicates a particular eigenvalue of $A$, as defined in Eq.\ref{eq:Lambda1A} to Eq.\ref{eq:Lambda3A}, and $a_1=2.04297$, $a_2=2.89648$ and $a_3=1.05859$ are constants, numerically obtained by the method.  Each $\lambda_j^{H}(x)$ follows these asymptotes, changing allegiance every time they intersect. There are also three intersection points, in increasing order of $x^\star_{12}\geq x^\star_{13}\geq x^\star_{23}$, with numerical values shown in Tab. \ref{tab:Intersections}.

\begin{table}[ht]
\centering
\begin{tabular}{ccc}
\toprule
\textbf{Intersection between} & $x^\star_{jk}$ & $y^\star_{jk}$ \\
\midrule
$f_1$ and $f_2$ & -0.22873 &  2.60814     \\
$f_1$ and $f_3$ & \ 0.17327 & 1.61484 \\
$f_2$ and $f_3$ & \ 0.94276 & 4.08503 \\
\toprule
\end{tabular}
\caption{{\bf  Intersections between asymptotes}. These are the  values of $x$ where each eigenvalue changes allegiance to one of the system's asymptotes. }\label{tab:Intersections}
\end{table}

\noindent The intersection are obtained by considering the $\lambda_k$ curves as hyperbolas, where their point of closest approach is where their derivatives are equal, which can be seen on Fig. \ref{fig:exDerivatives}. At last, it is possible to examine the three eigenvectors by taking their real spherical representation, i.e.,

\eq{
\theta_k = \cos^{-1}\left( U_{3k}^\star\ U_{3k} \right)\ \mbox{\hspace{1cm}and\hspace{1cm}}\ 
\phi_k = \tan^{-1}\left( U_{2k}^\star\ U_{2k} \right)\ ,
}

\noindent where it is implicit that $\sum_{jk} U_{jk}^\star\ U_{jk}=1$. Although this projection is a limited representation, where $V_k=(\theta_k,\phi_k )$, with $\theta/(\pi/2)\in[0,1] $ and $\phi/(\pi/4)\in[0,1]$, it is enough to observe the their limiting behavior, as shown on Fig. \ref{fig:exPolar}.

The initial position of each vector is $V_1=(0,1)$, $V_2=(1,1)$, and $V_3=(0,0)$, when $x=0$, revolving around the unit sphere for other values of $x$. When we compare each eigenvector with its corresponding eigenvalue on Fig. \ref{fig:exEigenvalues}, it is possible to correlate their behavior. For instance, how $V_1\rightarrow V_3$ as $x\rightarrow \pm\infty$ or how although $V_2$ crosses the other eigenvectors several times, the eigenvalues are never degenerate.

In conclusion, all the values and functions obtained in this example matches their analytical counterpart up to $10^{-14}$, which is the precision set for the method's precision $\varepsilon$, while the expected precision for evaluating the analytical solutions is $10^{-15}$ (using 64-bit floats).




\bibliography{references}

%



\end{document}